\documentclass[%
 reprint,
 amsmath,amssymb,
 prl,
longbibliography, 
]{revtex4-1}

\usepackage{graphicx}
\usepackage{dcolumn}
\usepackage{bm}
\usepackage{xcolor}

\begin{document}

\preprint{APS/123-QED}

\title{St. Petersburg paradox and failure probability}

\author{Jake Fontana}%
 \email{jake.fontana@nrl.navy.mil}
\affiliation{%
 U.S. Naval Research Laboratory\\
 4555 Overlook Ave. Southwest, Washington, DC 20375, USA
}%
\author{Peter Palffy-Muhoray}%
 \email{mpalffy@kent.edu}
\affiliation{%
 Advanced Materials and Liquid Crystal Institute\\
 Kent State University\\
 Kent, OH 44242, USA
}%

\date{\today}%

\begin{abstract}
The St. Petersburg paradox provides a simple paradigm for systems that show sensitivity to rare events.  Here, we demonstrate a physical realization of this paradox using tensile fracture, experimentally verifying for six decades of spatial and temporal data and two different materials that the fracture force depends logarithmically on the length of the fiber.  The St. Petersburg model may be useful in a variety fields where failure and reliability are critical. 


\end{abstract}

\maketitle


Failure of materials is ubiquitous.  In failure, breakdown often depends on the occurrence of rare events, such as large defects in mechanical \cite{RN1373} and dielectric \cite{RN1371,RN1370}  breakdown.  Although the Weibull model \cite{RN1281} has been used extensively over the last half century to describe such events, the model is empirically founded with no firm theoretical basis. It is of considerable interest then to further elucidate the relationship between material strength and scale \cite{RN1286, RN1284}.  

The St. Petersburg paradox \cite{RN1372} is a game in which the expectation value of winnings does not agree with the dictates of common sense.  A single trial in the St. Petersburg game consists of flipping a true coin until it lands heads; if this occurs on the nth flip, the payout is $2^n$ dollars.  The expectation value of the payout from a single trial is $\sum_{i=1}^{\infty} (1/2^{i}) 2^{i} = \infty$; however, in a typical trial, only a few dollars are won.  The paradox is that the expected outcome dramatically differs from the typical one.  The first full resolution of the paradox was given by Feller \cite{RN1877}. 

Given a sequence of $N$ coin tosses, i.e. $(10110,...,100101)$, where $1$ and $0$ correspond to the coin landing heads or tails, the mean number of clusters of length $n$ is $\left<m(n)\right>=(1-p)^{2}p^{n}N$ where $p$ is the probability of the coin landing tails. For large $N$, the system is expected to have many small clusters, and fewer large clusters.  No clusters of size $n$ are expected to occur with $ n> n_{max}$ where $(1-p)^{2}p^{n_{max}}N=0.5$.  The largest expected cluster size in a single chain is thus ${n_{max}}=-ln[2N(1-p)^{2}]/ln(p)$; that is, ${n_{max}}$ depends linearly on the logarithm of the chain length \cite{RN1370}.

\begin{figure}[htbp]
\centering\includegraphics[width=8.5cm]{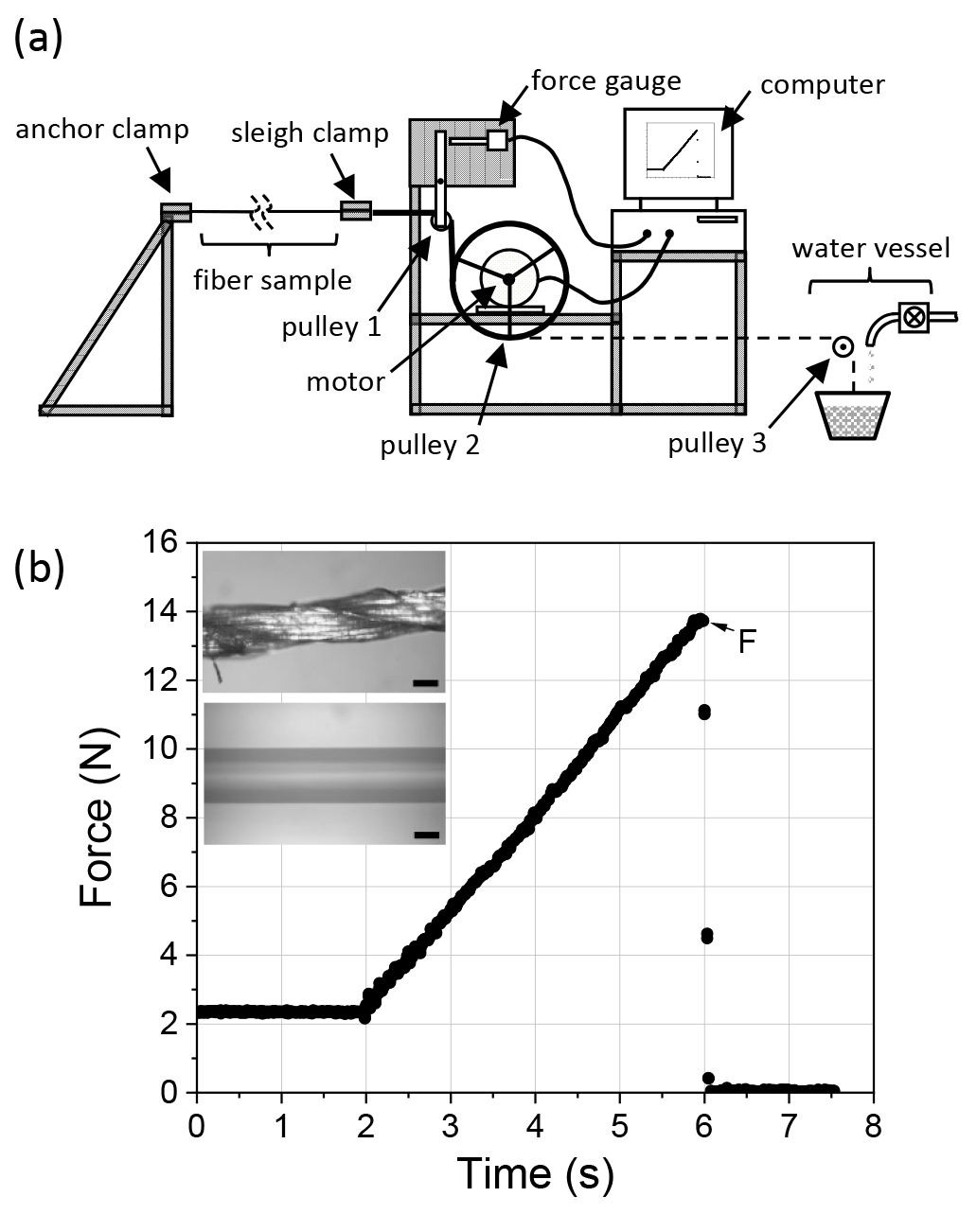}
\vspace{-0.5cm}
\caption{Schematic of the tensometer and force evolution on a fiber.  (a) The tensile load was applied and measured on the fiber samples as a function of length and time using a tensometer.  (b) A representative graph of the measured force as a function of time from a polyester fiber (Gutermann white sewing thread, filament diameter$=0.025~mm$, fiber diameter$=0.25~mm$).  The top left inset shows a microscope image of a polyester fiber sample and the bottom a polyamide fiber sample (Eagle Claw $6$ lbs Nylon monofilament fishing line, diameter$=0.22~mm$).  The scale bars in both images are equal to $0.1~mm$. }
\label{1}
\end{figure}

In this Letter, we report a physical realization of such behavior, by measuring the tensile force required to fracture fibers over six decades of fiber length and time for two materials.  Our results show that the force required to fracture a fiber depends linearly on the logarithm of the fiber length and is nearly independent of the strain rate. In addition to material failure, these results may have applications in fields such as weather forecasting, financial markets, internet congestion and hydrology. 

There is a fundamental connection between the St. Petersburg paradox and systems that show a sensitivity to rare events \cite{RN1283}.  In the St. Petersburg game, large profits result from the rare occurrence of long clusters of tails, i.e. $(1000,...,0001)$.  In failure, breakdown results from the rare occurrences of large defects.  If  ${n_{max}}$ is a measure of the size of the largest defect in a fiber length $L$ and if the force required to fracture the fiber is a linear function of the defect size, then the force $F$ required to fracture the fiber can be taken to depend linearly on the logarithm of the fiber length. Then

\begin{equation} 
\frac{F}{F_0}=-\alpha\ln\Big(\frac{L}{L_0}\Big)+\beta
\end{equation}

where $\alpha$ and $\beta$ are constants. $L_0$ and $F_0$ are normalizing constants and are set to the shortest fiber lengths that were measured in the experiments, described below, since the length and corresponding force required to fracture a defect-free fiber is initially unknown. 

\begin{figure}[htbp]
\centering\includegraphics[width=8.5cm]{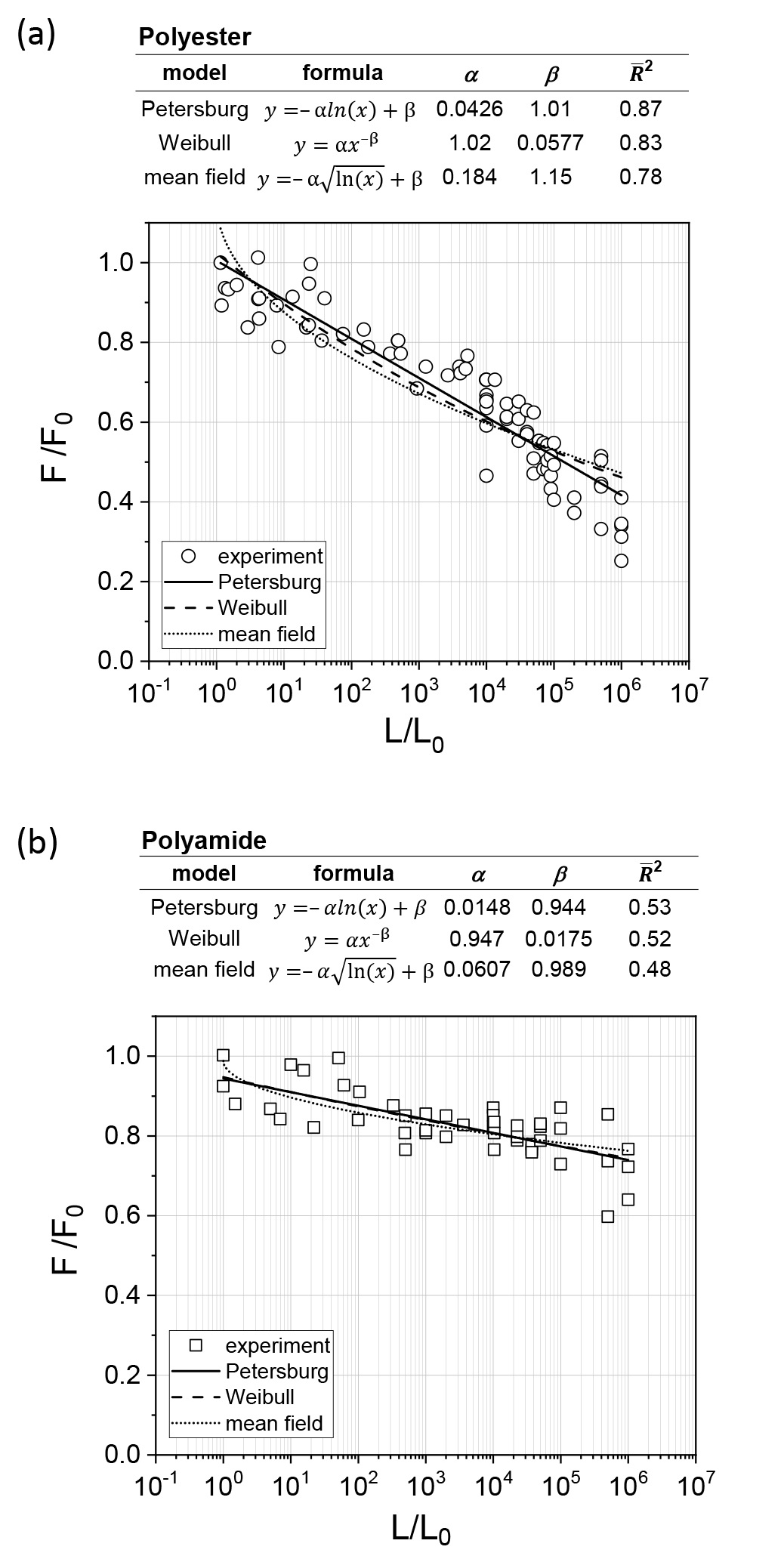}
\vspace{-0.5cm}
\caption{Normalized fracture force for different length fibers.  Normalized fracture force for the (a) polyester ($L_0=1.2~mm$, $F_0=17.84~N$) and (b) polyamide ($L_0=1.0~mm$, $F_0=42.60~N$) fibers as a function of fiber length from $1~mm$ to $1~km$.  The datasets were fit using the two parameter $(\alpha, \beta)$ St. Petersburg, Weibull and mean-field models and are summarized in the tables above each graph. $\bar R^{2}$ was used as a figure of merit. }
\label{1}
\end{figure}

To validate Eq. 1, we carried out tensile fracturing experiments on polyester and polyamide fiber samples with lengths ranging from $1~mm$ to $1~km$ (Fig. 1).

For fiber samples greater than $1~m$, the terminal end was clamped to a fixed anchor support, the other was laced over pulley $1$, attached to a lever that pushes on a force gauge (Omegadyne LC101-25), and is then anchored onto pulley $2$ (diameter, $24~cm$).  Pulley $2$ is attached to the shaft of an electric motor (Dayton $1/4$ hp, $323.5:1$, $0.095$ rev/s, ac gearmotor). Software (LabVIEW 8.0) using a data acquisition card (DAQ card PCI-6063E) and a relay controlled the motor, rotating pulley $2$, and measuring the applied load on the fiber from the force gauge until fracture occurred.  The fibers rarely fractured near the fiber ends; if this occurred, the data point was discarded.  The experiments were carried out on a straight $1~km$ paved bicycle trail during times of favorable weather.  Several supports were used to prevent the longer fibers from rubbing on the ground.

For fiber samples less than $1~m$, pulley $2$ was changed to a another with a diameter of $2~cm$ to reduce the strain rate, preventing any wave formation or propagation along the fibers, and was conducted indoors. The fibers were placed between two clamps, one fixed to the anchor support and the other to a sleigh clamp.  The sleigh clamp was connected to pulley $2$ using copper wire ($18$ gauge).  To prevent the fibers from slipping out of the clamps, epoxy drops were placed at the ends of the polyester fibers.  The polyamide fiber ends were tied (Snell knot) to the eyes of fishing hooks.  The epoxy beads or fishing hooks were then placed into both the sleigh and anchor clamps.  The strain rate of the tensometer was further reduced by detaching the copper wire from pulley $2$, lacing it over pulley $3$ and attaching it to a water vessel.  Using a valve the flow of the water into the vessel was controlled, from drops to full stream, thereby controlling the time required to fracture the fibers from days to seconds.  The time evolution of the force applied to a fiber is shown for a $5.3~m$ length polyester fiber [Fig. 1b].  Initially, a slight tension is applied to enable accurate measurements of the length of each fiber.  The electric motor is then started and tension builds until the fiber fractures.  

\begin{figure}[htbp]
\centering\includegraphics[width=8.5cm]{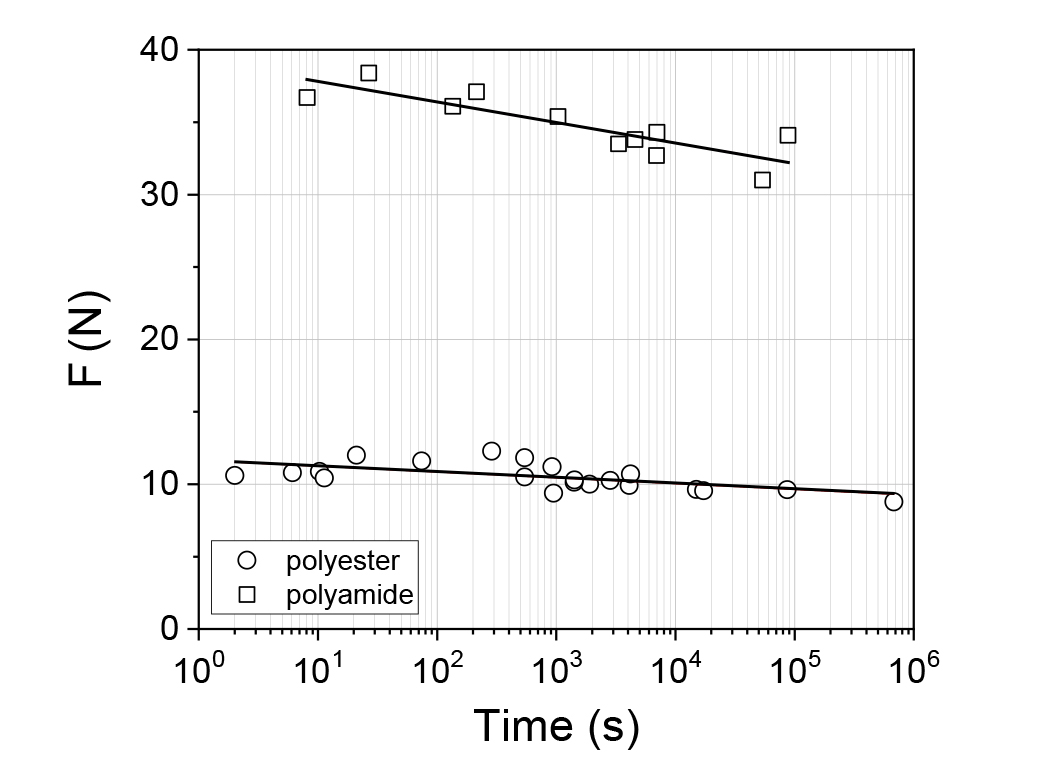}
\vspace{-0.5cm}
\caption{Fracture force for fibers at different strain rates. The force to fracture a polyester and polyamide fiber as a function of strain rate.  The data were fit using a linear regression.}
\label{1}
\end{figure}

The Weibull model asserts the probability, $P$ , that a single link will fracture at force $f$ is $P(f)=1-e^{-\phi(f)}$.  The cumulative probability of $n$ links failing can then be written as $P_n(f)=1-e^{-n\phi(f)}$.  Declaring  $\phi(f)$ of the form $((f-f_u)/f_0)^m$, where $f\geq f_u$  and $m,f_0>0$, and differentiating $P_n$ with respect to $f$ gives $P_w$, the Weibull distribution.  Letting $n$ be proportional to $L$ and maximizing $P_w$  with respect to $f$ gives the most probable force $F$ for link fracture, $F/F_0=\alpha(L/L_0)^\beta$ \cite{RN1281}.


To reveal how the fracture force depends on fiber length, the normalized fracture force for the polyester and polyamide fibers as a function of fiber length from $1~mm$ to $1~km$ are plotted (Fig. 2).  The datasets are fitted using the logarithmic St. Petersburg model [Eq. 1], the Weibull model, predicting a power law dependence, and a mean-field model \cite{RN1282}, $F/F_0=\alpha\sqrt{ln(L/L_0)}+\beta$, combining both a logarithmic and power law form.

The adjusted coefficient of determination, $\bar R^{2}$, was used as a figure of merit to compare the three models.  For both the polyester and polyamide datasets in Fig. 2, although the difference in the St. Petersburg and Weibull $\bar R^{2}$ is small, the St. Petersburg $\bar R^{2}$ was closest to unity, indicating that the logarithmic dependence of the fiber strength on length, predicted by the St. Petersburg model, agrees best with experimental results.

We note, properly the mean-field model should depend on three fitting parameters, since scaling $L_0$ in the $\sqrt{ln(L)-ln(L_0)}$ term does not simply scale $\alpha$ and $\beta$.  However, if the scaling factor of $L_0$ is close to unity, the scaling can be approximated by the change in $\alpha$ and $\beta$.  We have therefore included the mean-field model for the sake of completeness.

The length of a defect-free fiber can now be determined, $L_{f}=L_0e^{\frac{1-\beta}{\alpha}}$, from Eq. 1 when $\frac{F}{F_0}=1$. Using the fitting parameters retrieved from the tables in Fig. 2, we find for the polyester fiber, $L_{f}=1.5~mm$ and for the polyamide fiber, $L_{f}=0.022~mm$.

To probe how the strain rate affects the fiber fracturing force, the water vessel loading setup, discussed above, was used to vary the time required to fracture the fibers at a fixed length (Fig. 3).  The fracture force deviated from the average by $8.7\%$ and $6.2\%$ for the polyester ($L=6.2~m$) and polyamide ($L=3.8~m$) fibers, respectively, over a time span of $6$ orders of magnitude demonstrating a small, but non-negligible, creep dependence.  It is interesting to note that the dependence of the fracture force on strain rate also appears to be logarithmic \cite{Long2016}.

In conclusion, we demonstrated a physical realization of the St. Petersburg paradox using tensile fracture.  Our experiments show that the force required to fracture fibers with lengths from $1~mm$ to $1~km$ using two different materials, depends linearly on the logarithm of the fiber length.  These results indicate a fundamental connection between the St. Petersburg paradox and systems that show a sensitivity to rare events.

This work was supported in part by the Office of Naval Research (N0001420WX00146 and N00014-18-1-2624)

\bibliography{references}

\end{document}